# Bayes Blocks: An Implementation of the Variational Bayesian Building Blocks Framework

Markus Harva, Tapani Raiko, Antti Honkela, Harri Valpola, and Juha Karhunen
Neural Networks Research Centre
Helsinki University of Technology
P.O. Box 5400, FI-02015 TKK, Espoo, Finland
http://www.cis.hut.fi/projects/bayes/

## Abstract

A software library for constructing and learning probabilistic models is presented. The library offers a set of building blocks from which a large variety of static and dynamic models can be built. These include hierarchical models for variances of other variables and many nonlinear models. The underlying variational Bayesian machinery, providing for fast and robust estimation but being mathematically rather involved, is almost completely hidden from the user thus making it very easy to use the library. The building blocks include Gaussian, rectified Gaussian and mixture-of-Gaussians variables and computational nodes which can be combined rather freely.

## 1 INTRODUCTION

Variational Bayesian learning has been successfully used by many authors for solving a variety of learning problems (see e.g. Barber and Bishop, 1998; Attias, 1999; Ghahramani and Hinton, 2000; Valpola and Karhunen, 2002). The benefit of the variational approach is that it does not suffer as seriously from the problems that point estimation has, such as infinite densities, and yet it is computationally less demanding than sampling. In this sense the variational Bayesian learning combines the best qualities from both point estimation and sampling. However, deriving the necessary cost function and the update equations is a rather involved operation making variational Bayes less appealing.

The variational Bayesian building blocks framework, introduced by Valpola et al. (2001), offers a set of building blocks from which a large variety of different models can be constructed simply by connecting the blocks appropriately. This makes the construction of a model fast and easy as the framework automatically derives the cost function and the update rules, hiding all the hairy details from the user of the system.

We have written an efficient C++ implementation of the framework. The software package is called Bayes Blocks (Valpola et al., 2003a). For the convenience of use there is also a Python interface to the library providing easy scriptability with elegant syntax, fast development cycle as well as compact and easily readable code.

Works most closely related to ours are the Bayes Net Toolbox by Murphy (2001) and the VIBES software package by Bishop et al. (2003). The Bayes Net Toolbox can be used for Bayesian learning and inference of many types of directed graphical models using several methods. Hence it is in this sense more general than our library. But a serious limitation of the Bayes Net Toolbox is that it only supports latent continuous nodes with Gaussian or conditional Gaussian distributions. The reason is that more general models cannot be solved exactly.

The VIBES package implements the variational message passing (VMP) framework introduced by Winn and Bishop (2005). The VMP framework resembles the building block framework in that it uses variational Bayesian learning and factorised approximations. VIBES supports modelling some posterior dependencies whereas at the moment Bayes Blocks uses a fully factorial posterior approximation. The VIBES/VMP method differs from ours in that it mostly uses conjugate exponential family models (Gelman et al., 1995). Their Gaussian node uses a Gamma distribution for the precision parameter, which does not allow variance modelling in the same sense as the building block framework. Additionally, our building block framework supports nonlinearities.

A general drawback of the related methods by Murphy (2001) as well as Winn and Bishop (2005) is that they concentrate mainly on situations where there is a handy conjugate prior (Gelman et al., 1995) for the likelihood available. This makes life easier, but on the other hand our blocks can be combined more freely, allowing one to solve more difficult problems. The price we have to pay for this is that the minimum of the cost function for updating an individual node must sometimes be found iteratively, while it can be solved analytically when conjugate distributions are applied. The cost function can always be evaluated analytically in the building block framework as well. Our method could also be easily extended to handle the standard conjugate models.

We have applied the blocks mainly to different kinds of factor models. In Valpola et al. (2003b) hierarchical nonlinear factor analysis (HNFA) based on blocks was considered. In HNFA, the ability to learn the structure of the model is essential. Hierarchical modelling of variances was introduced by Valpola et al. (2004). There the application was to find variance sources from MEG data. The handling of missing and partially observed values by the use of evidence nodes was dealt with by Raiko (2004). The rectification nonlinearity is considered by Harva and Kabán (2005) who used it to perform non-negative factor analysis. Its application to astronomical data analysis is reported by Nolan et al. (2005).

The exposition of this paper proceeds as follows. In Section 2 a brief review of the most important aspects of the building blocks framework is given. A detailed description appears in a forthcoming paper (Raiko et al., 2005) the best references currently being the M.Sc. theses of two of the authors (Raiko, 2001; Harva, 2004). Some issues related to learning procedures and pruning are also discussed. Section 3 deals with the software library describing the implementation tools and the design philosophy. In Section 4 we report experimental results with a model structure that highlights some of the important points of the framework. Finally we have discussion and conclusions.

## 2 BUILDING BLOCKS FOR VARIATIONAL BAYESIAN LEARNING

In this section we give a brief review of the theoretical machinery underlying the library. The available building blocks are also introduced and combining them is discussed.

### 2.1 VARIATIONAL BAYES

In Bayesian data analysis, the data is usually assumed to have been produced by a generative model. The model typically has lots of uncertain quantities whose joint probability density function is to be estimated, given the data sample. This posterior density function contains all the relevant information on the unknown variables and parameters, but unfortunately it is often too complicated to handle without resorting to approximations. The simplest approximation is to consider only a single point of the distribution. This corresponds to maximum *a posteriori* or maximum likelihood approach. Unfortunately they suffer badly from overfitting especially when estimating the mean and variance of a parameter simultaneously. A second option is to draw samples from the posterior distribution, but in case of unsupervised problems, the number of unknown variables is so large that the computational complexity is often prohibitive.

Ensemble learning is a type of variational Bayesian learning, which employs the Kullback-Leibler divergence (information) between the approximate $q(\boldsymbol{\theta})$ and true posterior $p(\boldsymbol{\theta}|\boldsymbol{X})$, defined as

$$D\big(q(\boldsymbol{\theta})\big\|p(\boldsymbol{\theta}|\boldsymbol{X})\big) = \int q(\boldsymbol{\theta}) \log \frac{q(\boldsymbol{\theta})}{p(\boldsymbol{\theta}|\boldsymbol{X})} d\boldsymbol{\theta}.$$

The actual cost function is

$$\mathcal{C} = D\big(q(\boldsymbol{\theta})\big\|p(\boldsymbol{\theta}|\boldsymbol{X})\big) - \log p(\boldsymbol{X})$$

so that the evaluation of the problematic marginal likelihood term $p(\boldsymbol{X})$ is avoided. This cost function measures the misfit between the two distributions. Variational approximation is sensitive to probability mass rather than density and thus avoids the worst problems with overfitting. On the other hand, given a simple enough approximation $q(\boldsymbol{\theta})$, it is computationally efficient. Here, $q(\boldsymbol{\theta})$ is restricted to be fully factorial. This is one of the requirements for achieving a computational complexity that is linear with respect to the number of elements in the model. The method is also known as the naïve mean field approximation.

### 2.2 PROPAGATION OF INFORMATION

The building blocks can be divided into variable nodes and computational nodes. We shall refer to the connections to the parents and descendants of the nodes in the Bayesian network as their inputs and outputs, respectively (Cowell, 1999). Since the network is probabilistic, the values propagated between the nodes have associated distributions. When variational Bayesian learning together with a factorial posterior approximation is used, the cost function can be computed by propagating sufficient statistics instead of full distributions or functions. In the forward direction, the sufficient statistics are certain expectations (expected value, variance, and the expected exponential), and in the backward direction, similar statistics of the likelihood potential are propagated. These values carry all the information to adjust each variable to minimise the cost function (assuming other variables fixed). The minimisation can be done with a single step in some cases and iteratively in others.

## 2.3 VARIABLE NODES

Each variable node corresponds to a random variable, and it can be either observed or hidden. The most important type of a variable node is the Gaussian node. We also have the mixture-of-Gaussians node and the rectified Gaussian node. For a variable node, its inputs correspond to parameters of the conditional distribution of the variable represented by that node, and its output is the value of the variable.

Gaussian variable $s$ is conditioned by a mean $m$ and a variance $e^{-v}$

$$p(s|m, v) = \mathcal{N}(s|m, e^{-v})$$
$$= \frac{1}{\sqrt{2\pi e^{-v}}} \exp\left\{-\frac{1}{2e^{-v}}(s-m)^2\right\}.$$

The chosen parametrisation allows for the Gaussian node to be used as an input to both the mean and variance of another Gaussian node, which makes it useful in constructing hierarchical models.

Rectified Gaussian node takes the positive part of a zero-mean Gaussian

$$p(s|v) = \mathcal{N}^R(s|0, e^{-v}) = 2H(s)\mathcal{N}(s|0, e^{-v}),$$

where $H(s)$ is the Heaviside (a.k.a. the unit) step function. The rectified Gaussian node is useful if something known to be non-negative, such as energies, is modelled.

In the mixture-of-Gaussians node, a categorical variable $k \in \{1, \ldots, K\}$ selects one of the $K$ Gaussian components $\{\mathcal{N}(s|m_i, v_i)\}_{i=1}^{K}$

$$p(s|\{m_i\}_{i=1}^{K}, \{v_i\}_{i=1}^{K}, k) = \mathcal{N}(s|m_k, v_k)$$

The discrete variable $k$ has a Dirichlet prior. Mixture-of-Gaussians is used in e.g. independent factor analysis (Attias, 1999) as a source model.

## 2.4 COMPUTATIONAL NODES

For computational nodes, the output is a fixed deterministic function of the inputs. The computational nodes include the addition node, the multiplication node, a nonlinearity following a Gaussian node, and the delay node.

Addition node gives the sum of its inputs $\{s_i\}_{i=1}^{K}$ as the output $s_o$

$$s_o = \sum_{i=1}^{K} s_i.$$

Multiplication node produces the product of its inputs

$$s_o = s_1 s_2.$$

Together these two nodes can be used to form linear mappings required in almost any interesting model.

The blocks include two different nonlinearities

$$s_o = f(s) = \exp(-s^2),$$
$$s_o = f(s) = \max(s, 0).$$

The nonlinearities break the functional form of how the parent $s$ affects the cost function of its descendants. Therefore it is required that the nonlinearity is directly after a Gaussian variable. Nonlinearities are needed, e.g., for multi-layer perceptron (MLP) and radial basis function (RBF) network like structures.

The delay operation can be used to model dynamics. The node operates on time-dependent signals. It transforms the input $s(1), s(2), \ldots, s(T)$ into an output $s_0, s(1), s(2), \ldots, s(T-1)$ where $s_0$ is a scalar parameter that provides an initial distribution for the dynamical process.

## 2.5 COMBINING THE BLOCKS

Although the blocks can be combined rather freely there are some restrictions. Firstly, the network has to be a directed acyclic graph, like all Bayesian networks (Cowell, 1999). The delay nodes are an exception: the past values of any node can be the parents of any other nodes. Secondly, there should be only one computational path from a latent variable to another variable and different inputs of a node should be independent under the assumed posterior approximation. Thirdly, not all node types can be used as all types of parents of a node. These restrictions are summarised in Table 1. As an example, all nodes can serve as the mean of a Gaussian variable but only another Gaussian and summation can be used as the variance parent. This means that one cannot use a product node as the variance parent of a Gaussian variable, even if there is summation node in between. These restrictions can be evaded by adding intermediate variable nodes.

## 2.6 LEARNING PROCEDURE

The learning procedure aims at minimising the cost function. Many models have local minima in the cost function landscape so choosing a good initialisation can make a big difference. The best initialisation depends on the application and it is therefore left open. The initialisation can most conveniently be done using the so called evidence node which provides gradually fading virtual likelihood for its parent.

Table 1: Allowed connectivity

| Node | Parent | Type |
|---|---|---|
| $\mathcal{N}(m, e^{-v})$ | $m$ | any |
|  | $v$ | $\mathcal{N}, +$ |
| $\mathcal{N}^R(0, e^{-v})$ | $v$ | $\mathcal{N}, +$ |
| $\sum \pi_i \mathcal{N}(m_i, e^{-v_i})$ | $m_i$ | any |
|  | $v_i$ | $\mathcal{N}, +$ |
| $\sum s_i$ | $s_i$ | any |
| $s_1 s_2$ | $s_i$ | any |
| $f(s)$ | $s$ | $\mathcal{N}$ |

The basic element for updating the network is the update of a single node assuming the rest of the network fixed. One sweep of updating means updating each node once. The order in which this is done is not critical for the system to work. We have used an ordering where each variable node is updated only after all of its descendants have been updated. The progress can be measured by observing the monotonic decrease of the cost function.

The minimisation process can be easily accelerated by using pattern searches. The basic idea is that after some time, the parameters $\boldsymbol{\xi}$ describing $q(\boldsymbol{\theta})$ are strongly coupled and a single parameter $\xi_i$ can only be changed very little without changing the others as well. Now by collecting the individual updates $\Delta \xi_i$ from one full sweep, a line search can be performed to the direction of $\Delta \boldsymbol{\xi} = [\Delta \xi_1, \ldots, \Delta \xi_n]$. As there are both location and scale parameters among $\boldsymbol{\xi}$, some additional care needs to be taken in the line search procedure (Honkela et al., 2003).

## 3 THE SOFTWARE

The core functionality of Bayes Blocks is implemented in C++ with optional bindings to Python for scripting. The Python bindings are generated automatically with the help of SWIG (Beazley, 1996). Models are created by writing a Python script that defines the model structure. The system can then automatically derive the cost function related to the variational approximation and update rules for all the variables. The estimated models can be analysed using Python or saved in Matlab format. The latter option allows preparing visualisations of the results or analysing them further in Matlab.

### 3.1 LEARNING THE STRUCTURE OF THE MODEL

The Bayes Blocks software has been designed to deal with models whose structure changes during the learning process. Addition and removal of latent variables, weights and even larger hierarchical structures can be handled easily.

The basic operation for structural learning is checking whether the presence of an individual variable increases or decreases the overall cost function value. If the node in question affects its children through a linear mapping, the effect of the node's removal is equivalent to replacing it with a constant zero. Hence the effect on the children's cost function is easy to compute. The cost arising from the node itself is also readily available and consequently the net effect of removing the node is computable without changing the values of other variables. If the cost function is found to increase due to the variable, it will be pruned away. Information on the removal is propagated in the network and other nodes can act accordingly to adapt to the new situation. This means that removal of the weight $a_i$ in an inner product expression $a_1 s_1 + \cdots + a_n s_n$ correctly leads to removal of the product node for $a_i s_i$ and the corresponding term in the summation. If the present term is the only occurrence of $s_i$, it may be removed as well. The behaviours of different nodes in these situation can also be defined individually by the user.

Addition of new variables and connections to the network can be handled in the same manner as the initial creation of the network. The pruning method is useful here as it allows evaluation of the usefulness of the added parts and removal of unnecessary additions.

Since the pruning and addition of nodes are done based on the cost function which in its turn reflects the marginal likelihood of the model, the structural learning automatically implements Occam's razor. In other words, a simpler structure is favoured to a complex one even if the complex one would describe the data slightly more accurately.

### 3.2 MULTIDIMENSIONAL MODELS

Because of the emphasis on learning of the structure of the model, the handling of models for vectorial data in Bayes Blocks is somewhat atypical. Each component of a vector or a matrix is handled as a separate node which makes the structural learning operations easier. The only internally vectorised entities correspond to having multiple samples from a single scalar random variable. Because of their usage, all the vectors are currently assumed to have the same length.

## 4 EXPERIMENT

In this section an experiment with a dynamical model for variances applied to image sequence analysis is re-ported. The purpose is to demonstrate a nontrivial model implementable using Bayes Blocks.

The motivation behind modelling variances is that in many natural signals, there exists higher order dependencies which are well characterised by correlated variances of the signals (Parra et al., 2001). Hence we postulate that we should be able to better catch the dynamics of a video sequence by modelling the variances of the features instead of the features themselves. This indeed is the case as will be shown.

### 4.1 THE MODEL

The model considered can be summarised by the following set of equations:

$$\mathbf{x}(t) \sim \mathcal{N}(\mathbf{A}\mathbf{s}(t), \mathrm{diag}(\exp[-\mathbf{v}_x]))$$
$$\mathbf{s}(t) \sim \mathcal{N}(\mathbf{s}(t-1), \mathrm{diag}(\exp[-\mathbf{u}(t)]))$$
$$\mathbf{u}(t) \sim \mathcal{N}(\mathbf{B}\mathbf{u}(t-1), \mathrm{diag}(\exp[-\mathbf{v}_u]))$$

We will use the acronym DynVar to refer to this model. The linear mapping $\mathbf{A}$ from sources $\mathbf{s}(t)$ to observations $\mathbf{x}(t)$ is constrained to be sparse by assigning each source a circular region on the image patch outside of which no connections are allowed. These regions are still highly overlapping. The variances $\mathbf{u}(t)$ of the innovation process of the sources have a linear dynamical model. It should be noted that modelling the variances of the sources in this manner is impossible if one is restricted to use conjugate priors.

The sparsity of $\mathbf{A}$ is crucial as the computational complexity of the learning algorithm depends on the number of connections from $\mathbf{s}(t)$ to $\mathbf{x}(t)$. The same goal could have been reached with a different kind of approach as well. Instead of constraining the mapping to be sparse from the very beginning of learning it could have been allowed to be full for a number of iterations and only after that pruned based on the cost function as explained in Section 3.1. But as the basis for image sequences tend to get sparse anyway, it is a waste of computational resources to wait while most of the weights in the linear mapping tend to zero.

For comparison purposes, we postulate another model where the dynamical relations are sought directly between the sources leading to the following model equations:

$$\mathbf{x}(t) \sim \mathcal{N}(\mathbf{A}\mathbf{s}(t), \mathrm{diag}(\exp[-\mathbf{v}_x]))$$
$$\mathbf{s}(t) \sim \mathcal{N}(\mathbf{B}\mathbf{s}(t-1), \mathrm{diag}(\exp[-\mathbf{u}(t)]))$$
$$\mathbf{u}(t) \sim \mathcal{N}(\boldsymbol{\mu}_u, \mathrm{diag}(\exp[-\mathbf{v}_u]))$$

The variances $\mathbf{u}(t)$ are still estimated as part of the model but there is no longer a dynamical relation between them. We shall refer to this model as DynSrc.

### 4.2 IMPLEMENTATION USING BAYES BLOCKS

A possible Python implementation of the DynVar model is shown in Listing 1. Mostly the mathematical description of the model maps directly to corresponding statements in the implementation, although models are most conveniently constructed in a top-down fashion, as has been done here, too.

At line four a helper routine for constructing linear mappings from sum and product nodes is defined. It accepts an optional mask parameter to allow initially sparse mappings. This option is used at line 51, where the linear mapping from sources to observations is constructed. There the mask is as described in Section 4.1.

All nodes and variables are associated with a net, here constructed at line 26. Nodes are created through a node factory which conveniently allows for polymorphic constructors (Gamma et al., 1995). That is, we can provide an alternative implementation of the default factory to obtain nodes with modified implementation. There are two flavours of each node, one scalar and one vectorised, the length of the vectors being that given in the constructor of the net. Scalar and vector nodes can be combined in a natural manner—a vector node can have both scalar and vector parents whereas having a vector parent for a scalar node is not meaningful.

Since we require that all nets are acyclic, special care is needed with the use of delays since they create apparent cycles. To advice the library of this, so called proxy nodes are employed (at lines 35 and 46). They are given the label of the parent as their argument. At the end of the model construction, at line 59, the proxies are connected with a call to the net.

### 4.3 RESULTS

As the data, $\mathbf{x}(t)$, a video image sequence of dimensions $16 \times 16 \times 4000$ was used. That is, the data consisted of 4000 subsequent digital images of the size $16 \times 16$. A part of the dataset is shown in Figure 1.

We learnt both the models by iterating the learning algorithm 2000 times at which stage a sufficient convergence was attained. The first hint of the superiority of the DynVar model was offered by the difference of the cost between the models which was 28 bits/frame (for the coding interpretation, see Honkela and Valpola, 2004). To further evaluate the performance of the models, we considered a simple prediction task where the next frame was predicted based on the previous ones. The predictive distributions, $p(\mathbf{x}(t+1)|\mathbf{X}_{1:t})$, for the models can be approximately computed based on the posterior approximation. The means of the predic-

Listing 1: Variance model. See Sec. 4.2 in the text for details.

```
# Utility function for construction
# of linear mappings

def linmap(f, inputs, outdim, mask=None):
    sums = []
    a = []
    for i in range(outdim):
        sum = f.GetSumNV("sum")
        a.append([])
        for j in range(len(inputs)):
            if (mask is None) or mask[i,j]:
                a[i].append(
                    f.GetGaussian(
                        "a", c0, c0))
                p = f.GetProdV(
                    "prod", a[i][j],
                    inputs[j])
                sum.AddParent(p)
            else:
                a[i].append(None)
        sums.append(sum)
    return sums, a

# The core model

net = PyNet(tdim)
f = PyNodeFactory(net)

c0 = f.GetConstant("const0", 0.0)
cn5 = f.GetConstant("constneg5", -5.0)

pu = []
for j in range(sdim):
    pu.append(
        f.GetProxy("pu", Label("u", j)))

Bout, B = linmap(f, pu, sdim)

s = []
for j in range(sdim):
    vu = f.GetGaussian("vu", c0, cn5)
    du = f.GetDelayV("du", c0, Bout[j])
    u = f.GetGaussianV(
        Label("u", j), du, vu)

    ps = f.GetProxy("ps", Label("s", j))
    ds = f.GetDelayV("ds", c0, ps)
    s.append(f.GetGaussianV(
        Label("s", j), ds, u))

Aout, A = linmap(f, s, xdim, mask)

x = []
for i in range(xdim):
    vx = f.GetGaussian("vx", c0, cn5)
    x.append(
        f.GetGaussianV("x", Aout[i], vx))

net.ConnectProxies()
```

tive distributions are very similar for both of the models. Figure 2 shows the means of the DynVar model for the same sequence as in Figure 1. The means by themselves are not very interesting, since they mainly reflect the situation in the previous frame. However, the DynVar model also has a rich model for the variances. The standard deviations of its predictive distribution are shown in Figure 3. White stands for big variance and black for small. Clearly, the model is able to reduce the accuracy of the predictions in the area of high motion activity and hence provide better predictions. We can offer quantitative support for this claim by computing the predictive perplexities for the models. For each $t$ this is defined as

$$\exp\left\{-\frac{1}{256}\sum_{i=1}^{256} \log p(x_i(t+1)|\mathbf{X}_{1:t})\right\}.$$

The predictive perplexities for the same sequence as in Figure 1 are shown in Figure 4. Naturally the predictions get worse when there's movement in the video. However, DynVar model is able to handle it much better than its competitor.

## 5 DISCUSSION

The aims of the Bayes Blocks system are similar to the VIBES system by Bishop et al. (2003). The underlying frameworks and hence also the software implementations are, however, very different in practice.

The variational message passing (VMP) framework (Winn and Bishop, 2005) underlying the VIBES system is based on exponential family conjugate models. This excludes the hierarchical variance models as one cannot define a hierarchical model for the parameters of the Gamma distributed precision parameter of a Gaussian distribution. In the building block framework the precision of a Gaussian is log-normal and hence a hierarchical model can be easily defined recursively.

From software point of view, VIBES concentrates on building an easy-to-use graphical user interface with limited possibilities for the user. The Bayes Blocks software has no graphical user interface for defining models. Instead, the user gets the full power of Python scripting language. This allows very versatile connectivity of the blocks e.g. for sparse linear mappings that are often needed for handling very large data sets. The system supports easy pruning of unused parts of the model and addition of new parts allowing the model to be constructed iteratively, one part at a time.

The current Bayes Blocks is based on fully factorial posterior approximations. This design choice was made due to computational efficiency, as the pri-mary application for the library has been unsupervised learning of possibly rather large models. The library could of course be extended for modelling of posterior dependencies and using conjugate models.

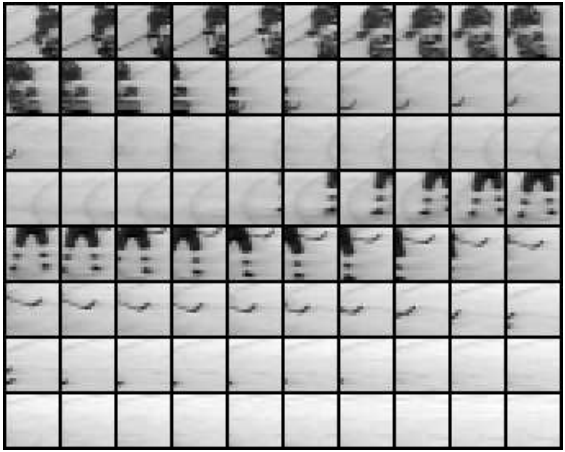

Figure 1: A sequence of 80 frames from the data used in the experiment.

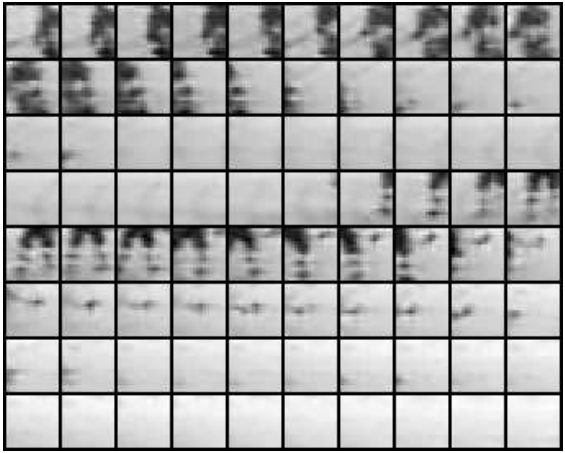

Figure 2: The means of the predictive distribution for the DynVar model.

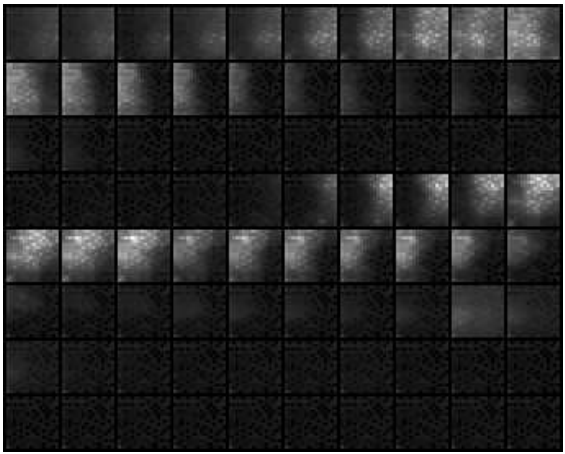

Figure 3: The standard deviations of the predictive distribution for the DynVar model.

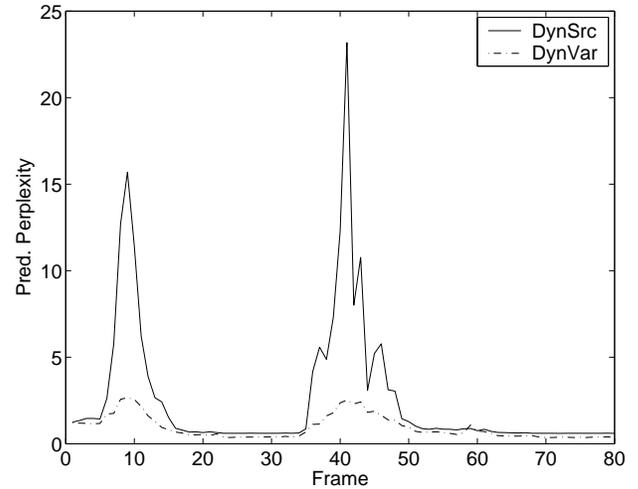

Figure 4: Predictive perplexities.

One of the advantages of Bayesian modelling is the ease of handling missing values (Raiko et al., 2003), or even partially missing values (Raiko, 2004). The latter is also known as virtual evidence and it means that a single value in data can be partially missing and partially observed. One might for example know that some value is probably greater than zero. Not only does the Bayes Blocks framework operate straightforwardly with these kind of inaccuracies in data, it also reconstructs the values during learning.

## 6 CONCLUSIONS

In this paper we presented the Bayes Blocks software library which is an implementation our variational Bayesian building blocks framework. It can be used to construct a rich class of probabilistic models by connecting simple variable and computational nodes appropriately. Underlying the library, there is the variational Bayesian methodology, which has proven to be a fast and reliable estimation scheme even for complex models. We showed by example how easily a nontrivial real-world model can be built using the library.

The library is free software and it is available for download at `http://www.cis.hut.fi/projects/bayes/software/`.

**Acknowledgements**

This work was supported in part by the IST Programme of the European Community, under the PAS-

CAL Network of Excellence, IST-2002-506778. This publication only reflects the authors' views. We would like to thank Hans van Hateren for supplying the video data (van Hateren and Ruderman, 1998).